# Tuning Superconductivity by Isovalent Antimony Substitution in PrFeAs(O,F)

Priya Singh[1], Konrad Kwatek[2], Tatiana Zajarniuk[3], Taras Palasyuk[2], Cezariusz Jastrzębski[2], Michał Wierzbicki[2], Tomasz Cetner[1], Andrzej Szewczyk[3], Shiv J. Singh[1]*

[1]*Institute of High Pressure Physics (IHPP), Polish Academy of Sciences, Sokołowska 29/37, 01-142 Warsaw, Poland*

[2]*Faculty of Physics, Warsaw University of Technology, Koszykowa 75, 00-662, Warsaw, Poland*

[3]*Institute of Physics, Polish Academy of Sciences, Aleja Lotnikow 32/46, 02-668 Warsaw, Poland*

***Corresponding author:**

Email: sjs@unipress.waw.pl

https://orcid.org/0000-0001-5769-1787

# Abstract

We investigate the effects of isovalent Sb substitution at the As site in fluorine-doped $PrFeAs_{1-x}Sb_xO_{0.7}F_{0.3}$ ($0 < x \leq 1.0$) through structural, Raman spectroscopy, density functional theory (DFT), transport, magnetotransport, and magnetic measurements. The superconducting transition temperature decreases gradually from ~48 K for the parent compound to ~44 K up to $x \approx 0.3$, followed by a rapid suppression at higher Sb concentrations due to increasing disorder and secondary phase formation. Raman spectroscopy and DFT reveal lattice expansion and pronounced softening of pnictogen-related vibrational modes upon Sb substitution. Magnetotransport measurements up to 9 T show enhanced upper critical fields ($H_{c2}$ ~ 200–350 T) and increased vortex activation energy for moderate Sb doping, indicating stronger vortex pinning. However, the critical current density remains low because of poor intergranular connectivity. The results demonstrate a crossover from an electronically tuned superconducting state to a disorder-dominated regime in isovalently substituted iron pnictides.

# I. Introduction

Iron-based superconductors (IBS) have established themselves as a prototypical platform for exploring unconventional superconductivity, where superconducting pairing emerges from a subtle interplay among electronic correlations, magnetism, and lattice degrees of freedom [1], [2]. Since the seminal discovery of superconductivity at $T_c = 26$ K in electron-doped $LaFeAsO_{1-x}F_x$, the so-called 1111 family (*RE*FeAsO, *RE* = La, Sm, Nd, Pr, Gd) has attracted sustained interest due to its rich phase diagram [3] [4] and high superconducting transition temperatures. In their parent composition, these materials exhibit a stripe-type antiferromagnetic ground state accompanied by a tetragonal-to-orthorhombic structural transition around 140–150 K. Upon chemical substitution—whether electron doping (e.g., F or H at the O site, Cu at the Fe site), hole doping (e.g., Sr at the *RE* site or Mn at the Fe site), or isovalent substitution (e.g., P or Sb at the As site)—the magnetic order is suppressed and superconductivity emerges, highlighting the intimate coupling between magnetism and superconducting pairing [5], [6], [7], [8], [9], [10], [11], [12], [13], [14], [15]. A defining characteristic of IBS is their multiband electronic structure, consisting of hole-like Fermi surface pockets at the Brillouin zone center (Γ point) and electron-like pockets at the zone corners (M point). The nesting between these pockets enhances interband scattering, which, within the spin-fluctuation framework, stabilizes an extended s± pairing symmetry [1] [16]. Notably, both the topology and orbital character of these Fermi surface sheets are highly sensitive to carrier concentration and structural parameters, rendering superconductivity in these systems highly tunable [17], [18].

Among various tuning strategies, isovalent substitution has proven to be a powerful approach for modifying the electronic structure without introducing additional charge carriers. In the Ba-122 ($BaFe_2As_2$) family, for example, phosphorus substitution at the arsenic site effectively induces superconductivity via chemical pressure, suppressing magnetic order and giving rise to quantum critical behavior [19] [20]. This substitution reduces the pnictogen height ($h_{pn}$), alters the Fermi surface topology, and drives a crossover from fully gapped to nodal superconductivity, underscoring the strong sensitivity of pairing symmetry to structural parameters [21]. Similarly, in the 11 family (e.g., FeSe), isovalent substitution with Te or S significantly modifies lattice parameters [22] and electronic correlations [23], leading to notable changes in superconducting properties, nematicity, and Fermi surface reconstruction [24]. These examples demonstrate that isovalent substitution can substantially influence low-energy electronic states without altering the carrier density. In contrast, the 1111 family remains comparatively less explored with respect to pnictogen-site isovalent substitution. While P substitution in LaFeAsO has been shown to induce superconductivity through chemical pressure effects [25], studies involving heavier pnictogens such as antimony (Sb) are limited. This distinction is crucial, as Sb substitution leads to lattice expansion and is expected to increase the pnictogen height, thereby accessing a region of the electronic phase space that is fundamentally different from that achieved via P substitution. Previous studies on La-based 1111 systems suggest that pnictogen substitution influences superconductivity through modifications of the Fermi surface nesting conditions [25]. However, recent investigations on SmFeAs(O,F) indicate that Sb substitution suppresses superconductivity, likely due to enhanced disorder and scattering effects [12]. These contrasting observations underscore the complex and system-dependent role of Sb substitution in the 1111 family. From a theoretical standpoint, an increase in the pnictogen height $h_{pn}$ enhances the contribution of the Fe $d_{xy}$ orbital near the Fermi level and may stabilize hole-like Fermi surface pockets [26]. At the same time, lattice expansion reduces orbital overlap, leading to bandwidth narrowing and enhanced electronic correlations [27]. Within the framework of multiband s± superconductivity, these effects compete with impurity-induced scattering, which acts as an efficient pair-breaking mechanism [16]. As a result, Sb substitution introduces a delicate balance between electronic structure tuning and disorder-driven suppression of superconductivity. In this context, PrFeAsO-based systems belonging to the 1111 family provide a particularly compelling platform owing to the presence of localized 4f moments, which interact with the FeAs layers and introduce an additional magnetic degree of freedom absent in La-based 1111 compounds. Recent studies have established that fluorine substitution in the range of 30–35% places $PrFeAsO_{1-x}F_x$ in the optimally doped regime, exhibiting a superconducting transition temperature $T_c$ of approximately 48.3 K [28]. To the best of our knowledge, the effects of Sb substitution in fluorine-doped PrFeAs(O,F) have not been systematically investigated previously.

In the present work, we investigate the impact of isovalent Sb substitution at the As site in optimally fluorine-doped $PrFeAsO_{0.7}F_{0.3}$. A comprehensive experimental and theoretical approach, combining structural characterization, electrical transport, Raman spectroscopy, magnetotransport, magnetic measurements, and density functional theory (DFT) calculations, is employed to investigate the evolution of superconductivity and the underlying electronic structure. By systematically examining the series $PrFeAs_{1-x}Sb_xO_{0.7}F_{0.3}$ ($x$ = 0 to 1.0), we aim to disentangle the competing effects of lattice modification, electronic correlations, and disorder-induced scattering. This study seeks to clarify the role of isovalent Sb substitution in tuning superconductivity within a magnetically active rare-earth 1111 system.

## II. Experimental Section

Polycrystalline samples of $PrFeAs_{1-x}Sb_xO_{0.7}F_{0.3}$ ($x$ = 0.05, 0.15, 0.20, 0.30, 0.40, 0.60, 0.80, and 1.0) were synthesized using a conventional two-step solid-state reaction method. High-purity starting materials: Pr (3N), $PrF_3$ (3N), $Fe_2O_3$ (4N), Fe (4N), As (6N, lumps), and Sb (3N)—were weighed according to the desired stoichiometry. All preparation steps, including weighing and mixing, were carried out inside an argon-filled glove box with oxygen and moisture levels maintained below 1 ppm to minimize contamination and oxidation. The powders were thoroughly ground and homogenized using an agate mortar and pestle for approximately 10–15 minutes. The homogeneous mixtures were then compacted into pellets of approximately 11 mm diameter under an applied pressure of ~200 bar. Subsequent thermal treatments were performed following standard procedures reported in the literature; further details of the synthesis protocol can be found in Refs. [28], [29].

Phase purity and crystal structure were characterized by powder X-ray diffraction (XRD) using a PANalytical Empyrean diffractometer with Cu Kα radiation (λ = 1.5418 Å), operated at 40 kV and 35 mA. Diffraction patterns were recorded over a $2\theta$ range of 10°–90° with a step size of 0.013° and a counting time of 300 s per step. Phase identification and qualitative analysis were carried out using the PDF4+ database from the ICDD. Rietveld refinement of the XRD data was performed to extract structural parameters, including lattice constants, and phase fractions, for the Sb-substituted samples. The elemental composition and microstructural features were investigated using energy-dispersive X-ray spectroscopy (EDX) attached to a high-resolution field-emission scanning electron microscope (SEM). Magnetic measurements were conducted using a vibrating sample magnetometer (VSM) integrated within a Physical Property Measurement System (PPMS, Quantum Design). The temperature dependence of magnetization was measured under zero-field-cooled (ZFC) and field-cooled (FC) conditions in an applied magnetic field of 20 Oe over the temperature range of 5–60 K. Magnetic hysteresis loops were recorded at 5 K under applied fields up to 9 T. Electrical resistivity was measured using a standard four-probe technique on rectangular specimens. Electrical contacts were prepared using silver paste and copper wires to ensure good electrical connectivity. The temperature dependence of resistivity in zero magnetic field was measured over the range 5–300 K using either a closed-cycle refrigerator or the PPMS. Magnetotransport measurements for selected compositions were carried out in the PPMS under magnetic fields up to 9 T and temperatures down to 5 K. Raman spectroscopy measurements were performed using a Horiba Jobin Yvon LabRAM ARAMIS spectrometer equipped with a 632.8 nm He–Ne laser as the excitation source. The incident laser beam was focused onto the sample surface using a 100× objective lens (numerical aperture = 0.95), resulting in a spot size below 3 μm. The backscattered signal was collected through the same objective, dispersed using a 2400 lines/mm grating, and detected with a CCD detector in the spectral range of 95–300 $cm^{-1}$. To minimize laser-induced heating effects, the laser power was limited to approximately 140 μW, and each spectrum was recorded with an accumulation time of 300 s. Measurements were performed directly on the as-synthesized sample surfaces. Given the polycrystalline nature of the samples, spectra were collected from at least five different locations for each composition to ensure reproducibility. The Raman-active phonon modes associated with the PrFeAs(O,F) phase were identified by fitting the Raman spectra using Lorentzian functions.

## III. Results and Discussions

### (a) Structural Analysis

The crystal structure of $PrFeAs_{1-x}Sb_xO_{0.7}F_{0.3}$ was investigated by powder X-ray diffraction (XRD), and the representative diffraction patterns are shown in Figure 1(a). All compositions predominantly crystallize in the tetragonal *P4/nmm* structure, characteristic of the 1111 phase. The parent compound ($x$ = 0) contains minor secondary phases, including FeAs, PrOF, and PrAs. Upon Sb substitution, these arsenide-related impurities are largely suppressed; instead, new impurity phases such as elemental Fe and Fe–Sb binaries ($FeSb/FeSb_2$) emerge with increasing Sb content. Despite their presence, the overall fraction of these secondary phases remains significantly lower than that of the dominant 1111 phase for moderate Sb concentrations. At higher substitution levels ($x \geq 0.6$), a substantial increase in impurity phases, particularly $Pr_2SbO_2$, is observed (see Supplementary Figure S1(a)), which begins to compete with the formation of the superconducting tetragonal phase. This behavior suggests that Sb is effectively incorporated into the FeAs lattice up to approximately $x \approx 0.6$, beyond which excess Sb preferentially forms secondary phases rather than entering the lattice. To further quantify the evolution of the secondary phases, the phase fractions are estimated from the Rietveld refinements and are summarized in Supplementary Table ST1. The analysis confirms that the tetragonal PrFeAs(O,F) phase remains the dominant phase up to $x$ = 0.6. For low and intermediate Sb concentrations ($x \leq 0.4$), the total impurity fraction remains relatively small and the superconducting PrFeAs(O,F) phase clearly dominates, whereas for $x$ = 0.6 a significant increase in the FeSb/$FeSb_2$ fraction (~26%) is observed accompanied by a small amount of $Pr_2SbO_2$ (~2%). The increasing concentration of these secondary phases at higher Sb contents correlates well with the progressive deterioration of the superconducting and transport properties discussed below and ultimately with the disappearance of superconductivity at higher Sb concentrations. For the optimal doped parent composition $PrFeAsO_{0.7}F_{0.3}$ ($x$ = 0), Rietveld refinement yields lattice parameters of $a = b = 3.978(6)$ Å and $c = 8.607(2)$ Å, consistent with previous reports [28]. Rietveld refinements were carried out for all Sb-substituted samples to extract structural parameters. A representative refinement for $x$ = 0.15 is shown in Figure 1(b), while additional refinements for $x$ = 0.05 and 0.40 are provided in Supplementary Figures S1(b) and S1(c), respectively. A systematic increase in the lattice parameter $c$ is observed with increasing Sb substitution ($x$), as illustrated in Figure 1(c). For comparison, previously reported data for Sb-doped SmFeAsO are also included [12] in this figure, showing a similar trend. The observed lattice expansion is consistent with isovalent substitution of the larger Sb atoms at the As-site, which is expected to modify the pnictogen height and the local Fe–pnictogen bonding environment without introducing additional charge carriers [11] [12]. Correspondingly, the unit cell volume ($V$), shown in the inset of Figure 1(c), exhibits a monotonic increase with Sb content ($x$), further confirming the structural expansion induced by Sb incorporation. For higher Sb concentrations ($x$ = 0.8 and 1.0), the XRD patterns reveal a significant increase in impurity phases, primarily PrOF and Pr–Sb–O compounds, which hinder the stabilization of the main PrFeAs(O,F) phase. Due to the substantial presence of these secondary phases, reliable extraction of lattice parameters for these compositions was not feasible; therefore, these data are not included in the structural analysis of this study.

To assess the actual Sb incorporation level, energy-dispersive X-ray spectroscopy (EDX) analysis was performed on the polished surface of these samples. A comparison between nominal Sb content ($x$) and experimentally determined composition ($x_{act}$), shown in Supplementary Figure S2, demonstrates good agreement between the nominal and actual Sb contents up to $x \approx 0.6$. This observation supports the conclusion that Sb is successfully incorporated into the FeAs lattice within this compositional range, while higher nominal doping leads to phase segregation.

## (b) Raman Spectroscopy

Raman scattering measurements were carried out to investigate the evolution of lattice dynamics and elucidate the impact of Sb substitution on the vibrational properties of $PrFeAs_{1-x}Sb_xO_{0.7}F_{0.3}$. Owing to its sensitivity to local bonding environments and site-specific atomic substitutions, Raman spectroscopy provides valuable insights into lattice modifications induced by isovalent doping. Figure 2 presents the evolution of the Raman spectra as a function of Sb concentration ($X_{Sb}$). The lower spectrum in Figure 2(a) acquired for the pristine material $PrFeAsO_{0.7}F_{0.3}$, i.e., with no Sb doping, reveals three prominent phonon modes at approximately 162 $cm^{-1}$, 204 $cm^{-1}$, 211 $cm^{-1}$. These modes are assigned to the out-of-plane vibrations of Pr ($A_1g$), As ($A_1g$), and Fe ($B_1g$), respectively, in agreement with previous reports and symmetry analysis [28] [29] [30] [31] [32]. With increasing Sb substitution up to $x \approx 0.30$, all three Raman-active phonon modes exhibit a systematic shift toward lower frequencies (phonon softening). This trend reflects a progressive weakening of the restoring forces associated with lattice vibrations. For intermediate doping levels ($0.30 < x \leq 0.60$), the Pr- and Fe-related modes continue to soften at a nearly constant rate, whereas the pnictogen-related ($A_1g$) mode associated with the As/Sb sublattice displays a significantly enhanced softening. The observed behavior can be attributed primarily to lattice expansion induced by the substitution of larger Sb atoms in place of smaller As atoms, consistent with the XRD results discussed above. The increase in lattice parameters leads to reduced bond stiffness, thereby lowering phonon frequencies. In addition to this structural effect, the substantial mass difference between Sb and As plays a crucial role. Since Sb is nearly twice as heavy as As, vibrational modes involving Sb atoms are expected to shift to lower energies. Density functional theory (DFT) calculations as depicted in the inset of Figure 2, further support this interpretation, predicting a pronounced reduction in the $A_1g$ phonon frequency of the pnictogen layer from ~220 $cm^{-1}$ in PrFeAsO to ~140 $cm^{-1}$ in the hypothetical fully substituted PrFeSbO compound. These results indicate that the enhanced phonon softening observed at higher Sb concentrations originates from the combined effects of lattice expansion (size effect) and the larger atomic mass of Sb (mass effect). Interestingly, the evolution of the pnictogen-related mode suggests two distinct regimes of phonon renormalization. At low Sb concentrations ($x \leq 0.30$), the frequency shift appears to be dominated by lattice expansion, whereas at higher Sb doping levels ($x > 0.30$), both lattice expansion and mass effects contribute significantly, resulting in an accelerated phonon softening of the vibrational modes. For high Sb concentrations ($x > 0.60$), reliable Raman signatures of the $PrFeAs_{1-x}Sb_xO_{0.7}F_{0.3}$ phase could not be clearly resolved due to the emergence of strong luminescence in the backscattered signal. This enhanced background is likely associated with the increasing presence of secondary phases, such as PrOF and Pr–Sb–O compounds, as also evidenced by XRD analysis. These impurity phases obscure the intrinsic phonon response of the superconducting $PrFeAs_{1-x}Sb_xO_{0.7}F_{0.3}$ phase, thereby limiting further spectroscopic analysis in this compositional regime.

## (c) Transport Properties

The temperature dependence of electrical resistivity ($\rho$) at zero magnetic field for $PrFeAs_{1-x}Sb_xO_{0.7}F_{0.3}$ was investigated to understand the evolution of charge transport and superconductivity with Sb substitution. The normal-state resistivity over a wide temperature range for up to $x = 0.6$ is shown in Figure 3(a), while the low-temperature region highlighting the superconducting transition is presented in Figure 3(b). The normal-state resistivity behaviour for the sample $x = 0.8$ and 1.0 is shown in the supplementary Figure S3. The parent compound ($x = 0$) exhibits a nearly linear temperature dependence of resistivity from room temperature down to the superconducting transition, consistent with optimally doped 1111 systems. A sharp superconducting transition is observed below ~48 K. This linear behavior persists for low Sb substitution levels up to $x \leq 0.3$, indicating that moderate isovalent substitution does not significantly alter the dominant scattering mechanisms in the normal state. At higher Sb concentrations ($x \geq 0.4$), the resistivity behavior deviates from this linear trend. In particular, a broad hump-like feature emerges in the temperature range above the superconducting transition, becoming more pronounced for $x = 0.4$ and 0.6 (Figure 3(a)). This feature likely originates from increased disorder and the presence of secondary phases, such as $FeSb_2$, as identified in the structural analysis. These impurity phases can introduce additional scattering centers and disrupt the coherence of charge transport. All samples up to $x = 0.6$ exhibit superconductivity; however, the normal-state resistivity shows a non-monotonic dependence on Sb content. While the room-temperature resistivity initially remains relatively unchanged, it increases significantly for higher Sb substitution, reaching a maximum near $x = 0.6$. This trend reflects the combined effects of impurity phase formation and enhanced scattering at grain boundaries. At low Sb content, residual metallic Fe inclusions may contribute to improved conductivity, whereas at higher doping levels, less conductive Fe–Sb phases ($FeSb/FeSb_2$) dominate and increase resistive scattering. For $x \geq 0.6$, the formation of complex impurity phases such as Pr–Sb–O compounds further disrupt the connectivity of the conducting FeAs network. For heavily substituted compositions ($x > 0.6$), the temperature-dependent resistivity (see Supplementary Figure S3) exhibits non-metallic behavior over the entire measured range, and no superconducting transition is observed. This behavior is consistent with the significant phase segregation identified in XRD measurements, which suppresses the formation of the superconducting 1111 phase. Similar increases in the resistivity arising from impurity phase formation and grain-boundary scattering have been widely reported in polycrystalline iron pnictides [33].

The low temperature behaviour of these samples is shown in Figure 3(b), which provides further insights into impact of Sb on the superconductivity. In the low substitution regime ($x \leq 0.3$), the superconducting transition temperature shows only a modest reduction from ~ 48 K ($x = 0$) to ~46-44 K, indicating that superconductivity is relatively robust against weak isovalent substitution. This behavior is consistent with the multiband nature of iron-based superconductors, where isovalent doping primarily modifies the electronic structure, such as orbital character and Fermi surface topology [20], without significantly altering carrier concentration. Similar trends have been reported in P-substituted $BaFe_2As_2$, where superconductivity persists over a wide doping range with only gradual changes in $T_c$ [19]. These results show that moderate Fermi-surface tuning does not immediately destroy the pairing mechanism as long as the nesting conditions remain intact. However, a distinct regime emerges for intermediate Sb concentrations ($0.3 < x \leq 0.6$), where a rapid suppression of superconductivity is observed. The transition temperature decreases sharply to ~34 K for $x$ = 0.4 and further to ~28 K for $x = 0.6$. This pronounced reduction in $T_c$ suggests a breakdown of the delicate balance required for multiband superconductivity. The simultaneous increase in the normal state resistivity and the emergence of impurity phases indicate that disorder and scattering effects become dominant at these Sb concentrations. The rapid suppression of $T_c$ beyond $x \approx 0.3$ highlights the critical role of impurity scattering and structural disorder, even in the absence of significant carrier doping. This behavior is consistent with previous studies on Sb-substituted 1111 systems [11] [12], where enhanced disorder leads to pair-breaking effects. In La-based compounds, the substitution of P at As-sites has been reported to enhance or weakly affect $T_c$ [25], whereas in Sm-based 1111 compounds, a systematic suppression of superconductivity with increasing Sb content is reported, accompanied by enhanced disorder effects [12]. Unlike P substitution, which typically enhances the Fe–Pn–Fe bond angle and can stabilize superconductivity, Sb substitution results in lattice expansion and is expected to modify the Fe–Pn–Fe bond angle, thereby weakening the electronic conditions favorable for superconductivity. Within the theoretical framework proposed by Usui *et al.* [33] for isovalently substituted iron pnictides (i.e. 1111 system), the superconducting properties are governed by a balance between orbital matching and the density of states at the Fermi level. In the low-doping regime ($x \leq 0.3$), this balance is only weakly perturbed, leading to a modest decrease in $T_c$. However, for higher Sb concentrations ($0.3 < x \leq 0.6$), the combined effects of reduced orbital overlap, degradation of Fermi surface nesting, and enhanced impurity scattering lead to a rapid suppression of superconductivity.

Magnetotransport measurements are carried out to investigate the influence of Sb substitution on the superconducting properties and vortex dynamics of the selected $PrFeAs_{1-x}Sb_xF_{0.7}O_{0.3}$ samples. Representative data for compositions $x = 0.15$ and 0.30 are presented in Figure 4(a) and 4(b), respectively. These compositions were selected due to their relatively sharp superconducting transitions ($\Delta T \approx 3$ K; see Figure 7(b)), which allow for a reliable analysis of field-induced broadening. The temperature dependence of resistivity was measured under applied magnetic fields up to 9 T. With increasing magnetic field, the superconducting transition progressively broadens and shifts toward lower temperatures, reflecting enhanced vortex motion and dissipation. To quantitatively determine the upper critical field ($H_{c2}$) and irreversibility field ($H_{irr}$), we adopted a standard resistive criterion of 90% and 10% of the normal-state resistivity ($\rho_n$) at $T = T_c^{onset}$ [34], [28], [29], respectively. From these criteria, the slope of the upper critical field near $T_c^{onset}$, $\left.\frac{dHc_2}{dT}\right|_{T=T_c}$ is extracted. For the parent compound ($x = 0$), the slope $\left.\frac{dHc_2}{dT}\right|_{T=T_c}$ is approximately -6.3 T/K, which increases to -6.7 T/K for $x = 0.15$ and further to -11.5 T/K for $x = 0.30$. This systematic enhancement indicates a strengthening of the orbital limiting field with increasing Sb content. On the other hand, the slope of the irreversibility field, $\left.\frac{dHirr}{dT}\right|_{T=T_c}$ decreases from −1.3 T/K for $x = 0$ to −0.9 T/K and −1.1 T/K for $x = 0.15$ and 0.30, respectively, as shown in Figure 4(c). The irreversibility-field slopes change only moderately with Sb substitution (−0.9 T/K for $x = 0.15$ and −1.1 T/K for $x = 0.30$ compared with −1.3 T/K for the parent compound), indicating modifications in vortex dynamics rather than a straightforward enhancement or suppression of vortex pinning. This behavior should be considered together with the TAFF results discussed below, where the activation energy $U_0$ increases upon Sb substitution, suggesting that the influence of Sb-induced disorder on vortex behavior is more complex than a simple strengthening or weakening of pinning. Within the framework of Werthamer-Helfand-Hohenberg (WHH) model for single band superconductor, the zero-temperature upper critical field is estimated using the formula [35]: $H_{c_2}(0) = -0.693 \cdot \left.\frac{dHc_2}{dT}\right|_{T=T_c} . T_c$. Using this approach, $H_{c_2}(0)$ for the parent compound is found to be ~212 T, which remains nearly unchanged (~214 T) for $x = 0.15$ and increases significantly to ~350 T for $x = 0.30$, reflecting enhanced scattering in the dirty-limit regime. The substantial enhancement of $H_{c_2}(0)$ with Sb substitution indicates a crossover toward a disorder-dominated (dirty limit) regime. In IBS, such behavior is commonly associated with increased scattering, which reduces the electronic mean free path and, consequently, the superconducting coherence length, thereby enhancing the orbital limiting field ($H_c^{orb}(0)$).

The corresponding coherence length is estimated using the Ginzburg–Landau relation $\xi_{GL}(0) = \sqrt{\frac{\Phi_0}{2\pi\mu_0 H_{c2}(0)}}$ [36]. The calculated values are $\xi_{GL} \approx 1.24$ nm for $x = 0.15$ and $\approx 0.97$ nm for $x = 0.30$, compared to $\approx 1.23$ nm for the parent compound ($x = 0$) [28]. The noticeable reduction in coherence length for $x = 0.30$ further supports the presence of enhanced impurity scattering at higher Sb concentrations, while only marginal changes are observed at lower substitution levels ($x = 0.15$). It is important to note that deviations from the single-band WHH model are expected in this system due to its multiband nature. In particular, the large estimated Maki parameter $\alpha$ (= $\sqrt{2}\,\frac{H_c^{orb}(0)}{H_P}$) [37], which reflects the competition between orbital and Pauli paramagnetic limiting effects, suggests that spin-paramagnetic contributions may play a significant role. Here, $H_c^{orb}$ and $H_P$ represent the orbital and Pauli paramagnetic upper critical field, respectively. Such behavior has been widely reported in iron-based superconductors [28], [29] and highlights the limitations of applying a single-band WHH description to these complex systems. Overall, the magneto-transport results reveal that moderate Sb substitution enhances the upper critical field through increased scattering and modifies the vortex dynamics. When considered together with the TAFF results, the data suggest that Sb-induced disorder introduces additional pinning centers that enhance the activation energy for vortex motion, while simultaneously altering the irreversibility field and flux dynamics.

The thermally activated flux flow (TAFF) behaviour in Sb-doped $PrFeAs_{1-x}Sb_xF_{0.7}O_{0.3}$ provides an important insight into the dimensionality of vortex dynamics and pinning mechanisms, shown in Figures 5(a) and 5(b). The TAFF regime provides key insight into vortex motion in the presence of thermal fluctuations and disorder [38]. The activation energy $U_0$ associated with vortex motion is extracted from Arrhenius plots of resistivity in the flux flow regime using the relation: $\rho\,(T,H) = \rho_n\, exp\left(-\frac{U_0(H)}{k_B T}\right)$: where, $\rho_n$ is the normal-state resistivity prefactor, $k_B$ is the Boltzmann constant, and $U_0(H)$ is the effective activation energy of the vortex motion. The extracted activation energies for different compositions are summarized in Figure 5(c). For clarity, the Arrhenius plots for $x = 0.15$ and $x = 0.30$ are presented using identical $x$- and $y$-axis scales in Figures 5(a) and 5(b). Both compositions exhibit approximately linear behavior within the fitting region employed for the TAFF analysis, confirming the validity of the extracted activation energies. The activation energy $U_0$ associated with vortex motion exhibits a clear enhancement for both $x$ = 0.15 and $x$ = 0.30 compared with the parent compound ($x$ = 0), as shown in Figure 5(c). This increase in $U_0$ indicates an enhanced energy barrier for vortex motion and is consistent with improved vortex pinning in the Sb-substituted samples. Furthermore, the activation energy for $x$ = 0.30 remains comparable to or slightly higher than that for $x$ = 0.15 over most of the investigated field range, suggesting that the additional disorder introduced by Sb substitution acts as an effective source of pinning centers without significantly degrading the vortex pinning landscape.

The field dependence of the activation energy follows a power-law behavior, $U_0 \sim H^{-\eta}$, which reflects the nature of vortex pinning. For the parent compound ($x$ = 0) [28], two distinct regimes are observed, characterized by exponents $\eta_1 \approx 0.30$ in the low-field region and $\eta_2 \approx 0.69$ in the high-field region. This crossover signifies a transition from a single-vortex or weak pinning regime at low fields to a collective flux-flow regime at higher fields, where vortex–vortex interactions become significant. With moderate Sb substitution ($x$ = 0.15), the low-field exponent increases to $\eta_1 \approx 0.37$, while the high-field exponent remains nearly unchanged ($\eta_2 \approx 0.69$). This evolution suggests a slight enhancement of pinning strength in the single-vortex regime without significantly altering the collective behavior at higher fields. The increased activation energy and the suppression of double-step features in the magnetic transition (Figure 6(a)) further indicate reduced weak-link effects and improved intergranular coupling. For higher substitution ($x$ = 0.30), the exponents evolve to $\eta_1 \approx 0.49$ and $\eta_2 \approx 0.59$. The continued increase in $\eta_1$ reflects further strengthening of pinning in the low-field regime, likely due to increased disorder and the presence of Fe–Sb-related secondary phases acting as effective pinning centers. At the same time, the modification of the high-field exponent suggests the changes in vortex–vortex interactions and collective dynamics, possibly driven by enhanced scattering and structural inhomogeneity. In the parent compound, relatively weaker pinning and stronger vortex motion can be attributed to phase inhomogeneity and weak intergranular coupling. In contrast, Sb substitution introduces non-magnetic disorder and secondary phase inclusions, which act as additional pinning centers and enhance the overall pinning landscape. This is consistent with the observed increase in $U_0$ and the suppression of weak-link behavior. The observed two-regime field dependence of $U_0$ is consistent with collective pinning and creep models, where $\eta$ values in the range ~0.2–0.5 correspond to single-vortex or weak collective pinning, and higher values (~0.5–1) indicate collective creep dominated by vortex interactions [38] [39]. The systematic evolution of $\eta$ with Sb content demonstrates that isovalent substitution effectively tunes the vortex dynamics and pinning mechanisms in this system. It is instructive to contrast this behavior with that of Mn-substituted $PrFe_{1-x}Mn_xAsO_{0.7}F_{0.3}$ [29], where Mn acts as a magnetic dopant. In Mn-doped systems, although low doping levels may initially enhance $U_0$ due to additional pinning centers, higher Mn concentrations lead to a suppression of activation energy below that of the parent compound. This is typically attributed to strong magnetic pair-breaking effects and reduced superconducting coherence. In contrast, Sb substitution, being isovalent and non-magnetic, enhances the activation energy over the investigated compositional range, indicating that non-magnetic disorder can improve vortex pinning without introducing significant pair-breaking effects. Overall, the TAFF analysis highlights that Sb substitution not only modifies the electronic and structural properties but also plays a crucial role in optimizing vortex pinning, particularly in the low-field regime, while maintaining robust superconductivity at moderate doping levels.

## (d) Magnetic Properties

The magnetic response of $PrFeAs_{1-x}Sb_xO_{0.7}F_{0.3}$ ($x$ = 0, 0.05, 0.15, 0.30, 0.40 and 0.60) is investigated through zero-field-cooled (ZFC) and field-cooled (FC) magnetization measurements under an applied magnetic field of 20 Oe, as shown in Figure 6(a). The superconducting transition temperature $T_c^{mag}$ is determined from the onset of diamagnetism, defined by the bifurcation point between the ZFC and FC curves. The parent compound ($x$ = 0) exhibits a superconducting transition in magnetization at $T_c^{mag} \approx 46.1$ K. A low level of Sb substitution i.e., $x = 0.05$ results in a superconducting transition temperature nearly identical to that of the parent sample, as depicted in the supplementary Figure S4. With increasing Sb substitution, a systematic suppression of $T_c^{mag}$ is observed, decreasing to ~46 K ($x = 0.15$), ~42.5 K ($x = 0.30$), ~33 K ($x = 0.40$), and ~26 K ($x = 0.60$), as summarized in Figure 6(a). This trend is in good agreement with the superconducting transition temperatures obtained from resistivity measurements, confirming the progressive weakening of superconductivity with increasing Sb content. Notably, the magnetic transition temperatures $T_c^{mag}$ are consistently lower by approximately 1–2 K compared to those derived from transport measurements. Such a discrepancy is commonly observed in polycrystalline iron-based superconductors [40] and reflects the difference between the percolative nature of electrical transport and the requirement of bulk diamagnetic screening for magnetic measurements. While resistivity detects the onset of superconducting pathways, magnetization probes the establishment of a fully connected superconducting volume. A key observation is the evolution of the transition shape with Sb substitution. The parent compound ($x$ = 0) exhibits a characteristic two-step diamagnetic transition, indicative of weak-link behavior and superconducting inhomogeneity, typically associated with poor intergranular coupling. In contrast, Sb-substituted Pr1111 samples display a single, sharper diamagnetic transition, suggesting a more homogeneous superconducting response and reduced weak-link behaviour. This improvement in magnetic response correlates well with the thermally activated flux flow analysis. The enhanced activation energy ($U_0$) observed for $x$ = 0.15 and 0.30 indicates an increased energy barrier for vortex motion and reduced flux creep, which contribute to stabilizing the superconducting state. Nevertheless, as discussed below, the global current-carrying capability remains limited by secondary phases and imperfect grain connectivity. Therefore, the magnetic measurements suggest that moderate Sb substitution improves the local superconducting homogeneity and vortex pinning properties, even though the macroscopic intergranular current transport remains relatively weak. Hence, the magnetic measurements reinforce the transport and vortex dynamics results, demonstrating that while Sb substitution gradually suppresses $T_c$, it simultaneously improves vortex pinning and reduces intergranular weak-link effects at moderate doping levels.

Magnetic hysteresis ($M$–$H$) loops were measured for selected compositions to evaluate the current-carrying capability of $PrFeAs_{1-x}Sb_xF_{0.7}O_{0.3}$. The critical current density $J_c$ was estimated using the Bean critical state model, $J_c = 20\Delta m/[a(1 - a/3b)V]$, where $\Delta m$ is the width of the magnetization loop, $V$ is the sample volume, and $a$ and $b$ ($a < b$) are the sample dimensions. The field dependence of $J_c$ is presented in Figure 6(b). The Sb-substituted samples exhibit relatively low critical current densities, of the order of $10^2$ A cm$^{-2}$ or lower, which is significantly smaller than that of the parent compound ($x$ = 0) [28], where $J_c$ ~$10^3$ A cm$^{-2}$. This pronounced reduction indicates that Sb substitution adversely affects the global current-carrying capability of the polycrystalline samples. The suppression of $J_c$ can be primarily attributed to the degradation of intergranular connectivity induced by the evolution of secondary phases. As discussed in the transport and structural sections, the system undergoes a progression from Fe-rich impurity phases at low Sb content ($x \leq 0.3$) to Fe–Sb binaries (FeSb/$FeSb_2$), and eventually to more complex Pr–Sb–O phases at higher substitution levels. These non-superconducting inclusions disrupt the continuity of superconducting pathways across grain boundaries, thereby limiting long-range current percolation. In polycrystalline IBS, the global $J_c$ is strongly governed by intergranular coupling rather than intragrain pinning alone. Even in the presence of enhanced vortex pinning, as evidenced by the increased activation energy from TAFF analysis, the presence of weak links at grain boundaries can severely constrain the effective transport of supercurrents. Consequently, while Sb substitution appears to improve local pinning properties, the simultaneous deterioration of grain connectivity dominates the overall current-carrying performance. These results highlight the dual role of Sb-induced disorder: it enhances local vortex pinning and increases the activation energy for vortex motion, while simultaneously promoting secondary-phase formation and structural inhomogeneity that weaken intergrain coupling and suppress the macroscopic critical current density.

### (e) Discussion

Figure 7 summarizes the evolution of key superconducting and normal-state parameters for $PrFeAs_{1-x}Sb_xO_{0.7}F_{0.3}$ ($0 \leq x \leq 0.6$), including the onset superconducting transition temperature $T_c^{onset}$ from the transport measurements, transition width $\Delta T$, residual resistivity $\rho_0$ obtained from extrapolation of the normal-state resistivity to $T = 0$ K and residual resistivity ratio ($RRR$) defined by $RRR = \rho_{300K} / \rho_{50K}$ and $RRR = \rho_{300K} / \rho_0$. These parameters provide a comprehensive overview of the impact of isovalent Sb substitution on the electronic, structural, and microstructural properties of the Pr1111 system. A systematic suppression of $T_c^{onset}$ is observed with increasing Sb content, decreasing from ~48.3 K for $x = 0$ to ~28 K for $x = 0.6$ (Figure 7(a)). This trend reflects a gradual weakening of superconductivity and is consistent with previous reports on Sb-substituted 1111 systems, including $SmFeAs_{1-x}Sb_xO_{0.8}F_{0.2}$ compounds [12]. From a microscopic perspective, isovalent Sb substitution does not significantly change the carrier concentration but modifies the electronic structure through lattice expansion. As evidenced by the XRD results, the larger ionic radius of Sb compared with As leads to a systematic increase in the lattice parameters, particularly along the $c$-axis. Such lattice expansion is expected to modify the local Fe–Pn tetrahedral environment [41] and may increase the pnictogen height ($h_{pn}$) while simultaneously shifting the Fe–Pn–Fe bond angle away from the ideal tetrahedral value of 109.47° [26, 27, 41]. These structural parameters are known to play a crucial role in determining the electronic structure of IBS. Previous theoretical studies have shown that variations in $h_{pn}$ and the Fe–Pn–Fe bond angle strongly influence the orbital composition and topology of the Fermi surface [26, 27, 41]. In particular, an increase in $h_{pn}$ enhances the contribution of the Fe $d_{xy}$ orbital near the Fermi level and modifies the relative positions of the hole-like bands around the Γ point. Simultaneously, the increase in Fe–Pn bond length reduces Fe–Pn orbital overlap, leading to narrower electronic bandwidths and enhanced electronic correlations. Possible changes in the Fe–Pn–Fe bond angle are expected to modify the hopping amplitudes between neighbouring Fe atoms, thereby altering the size and orbital character of the hole pockets at Γ and the electron pockets at M. Consequently, the degree of electron–hole nesting, which is believed to play an important role in spin-fluctuation-mediated superconductivity in iron pnictides, may be progressively weakened. Within the theoretical framework proposed by Kuroki et al. and Usui et al. [26, 33, 41], superconductivity is highly sensitive to the pnictogen height and the orbital matching between electron and hole Fermi-surface sheets. Therefore, the relatively weak suppression of $T_c$ observed at low Sb concentrations ($x \leq 0.3$) suggests that the electronic structure remains close to the optimum configuration for superconductivity. In contrast, at higher Sb concentrations ($0.3 < x \leq 0.6$), the combined effects of lattice expansion, possible changes in $h_{pn}$ and Fe–Pn–Fe bond angle, and enhanced disorder-induced scattering are expected to progressively deteriorate the nesting conditions and orbital matching. This scenario provides a plausible explanation for the rapid suppression of $T_c$ observed experimentally.

In contrast to the monotonic suppression of $T_c^{onset}$, the transition width $\Delta T$ exhibits a non-monotonic evolution (Figure 7(b)). At low Sb concentrations ($x \leq 0.3$), $\Delta T$ decreases significantly from ~6 K for $x = 0$ to ~3 K for $x = 0.3$, indicating improved superconducting homogeneity and reduced weak-link behavior. This observation is consistent with magnetic measurements, where the disappearance of the double-step transition in the ZFC curves suggests a more homogeneous superconducting response and reduced weak-link behaviour. Furthermore, the increase in vortex activation energy ($U_0$) obtained from TAFF analysis supports the presence of stronger pinning and reduced flux creep in this regime. However, for higher Sb concentrations ($0.3 < x \leq 0.6$), $\Delta T$ increases markedly, indicating a deterioration of superconducting uniformity. This broadening correlates with the emergence of structural inhomogeneity, as evidenced by XRD and Raman measurements, which reveal increasing impurity phases and local lattice distortions. Such inhomogeneity leads to spatial variations in superconducting properties and promotes percolative superconductivity, where the transition is governed by the connectivity of superconducting regions rather than intrinsic bulk behavior. A similar non-monotonic evolution of $\Delta T$ has been reported in Sb-substituted Sm-1111 systems [12], where disorder effects dominate at higher doping levels.

The evolution of the residual resistivity ($\rho_0$), shown in Figure 7(c), provides further insight into the role of disorder introduced by Sb substitution. The residual resistivity $\rho_0$, obtained by extrapolating the normal-state resistivity to $T = 0$ K using the relation $\rho = \rho_0 + AT^n$, serves as a direct measure of impurity and defect scattering. Representative fits and the extrapolation procedure for selected compositions ($x$ = 0.05, 0.20, and 0.40) are presented in Supplementary Figure S5. As shown in Figure 7(c), $\rho_0$ remains relatively small for low Sb concentrations ($x \leq 0.2$), indicating that moderate Sb substitution introduces only limited additional scattering. However, $\rho_0$ increases rapidly for $x \geq 0.3$ and reaches approximately 5–6 mΩ-cm for $x = 0.6$, demonstrating a substantial increase in disorder-induced scattering. This trend is consistent with the increasing fraction of Fe–Sb and Pr–Sb–O secondary phases identified by the structural analysis. The sharp increase in $\rho_0$ above $x \approx 0.3$ coincides with the onset of rapid $T_c$ suppression and increasing impurity-phase content, suggesting that disorder scattering becomes one of the dominant factors governing the superconducting properties in this compositional regime.

For comparison, Figure 7(d) presents both the conventional residual resistivity ratio, $RRR = \rho_{300K} / \rho_{50K}$, and the modified ratio, $RRR = \rho_{300K} / \rho_0$. The latter definition is widely regarded as a more direct indicator of sample quality and disorder because it explicitly reflects the residual scattering contribution. A slight enhancement in the conventional *RRR* at $x = 0.15$ indicates improved metallicity, possibly arising from reduced impurity scattering (e.g., suppression of Fe-related inclusions) and improved local superconducting homogeneity. However, at higher Sb concentrations, both *RRR* values decrease systematically, reflecting enhanced impurity scattering and degradation of the normal-state electronic transport. In particular, the modified residual resistivity ratio decreases dramatically from approximately 33.7 for the parent compound to 1.85 for $x = 0.6$. This nearly eighteen-fold reduction provides compelling evidence for the rapid increase in disorder-induced scattering with increasing Sb substitution. Such behavior is consistent with the progressive formation of Fe–Sb and Pr–Sb–O secondary phases identified in the structural analysis and agrees with previous reports on Sb-doped Sm-1111 [12]. While the conventional *RRR* exhibits a gradual decrease with increasing Sb content, the ratio based on $\rho_0$ decreases much more strongly, highlighting the rapid growth of residual scattering at higher Sb concentrations. Similar correlations between increasing residual resistivity, decreasing *RRR*, and suppression of superconductivity have been reported in several iron-based superconductors [34], where enhanced disorder weakens superconducting pairing and degrades electronic transport [42]. The simultaneous increase of $\rho_0$ and decrease of both *RRR* values therefore demonstrate that disorder-induced scattering becomes increasingly important with Sb substitution and contributes significantly to the suppression of superconductivity for $x > 0.3$. Taken together, these results reveal a crossover from a relatively homogeneous superconducting state at low Sb substitution to a disorder-dominated regime at higher doping levels. At low Sb content ($x \leq 0.3$), moderate lattice expansion and controlled disorder lead to improved superconducting homogeneity, enhanced vortex pinning, and relatively robust superconductivity. On the other hand, at higher Sb doping levels ($0.3 < x \leq 0.6$), excessive disorder, phase segregation, and degradation of intergranular connectivity collectively suppress the bulk superconducting response and broaden the superconducting transition. The simultaneous evolution of $T_c$, $\Delta T$, $\rho_0$, and $RRR$ therefore suggests that the superconducting properties of $PrFeAs_{1-x}Sb_xO_{0.7}F_{0.3}$ are governed by a competition between electronic-structure modifications induced by lattice expansion and disorder-induced scattering associated with increasing Sb substitution

Importantly, the present study demonstrates a strong coupling between structural (XRD), vibrational (Raman), and electronic (transport and magnetotransport) responses in $PrFeAs_{1-x}Sb_xO_{0.7}F_{0.3}$. Compared with previously reported Sb-substituted Sm1111 systems [12] the Pr-based 1111 compound exhibits similar correlations between lattice expansion, phonon softening, vortex dynamics, and superconducting properties, while providing additional insight into the interplay between disorder and superconductivity in a magnetically active rare-earth system. This highlights the critical role of lattice–electronic coupling and disorder in governing superconductivity in isovalently substituted iron pnictides, and provides a unified framework for understanding the interplay between structure, electronic correlations, and vortex physics in these systems.

# IV. Conclusion

We systematically investigated the effects of isovalent Sb substitution at the As site in $PrFeAs_{1-x}Sb_xO_{0.7}F_{0.3}$ ($0 < x \leq 1.0$) using structural, Raman spectroscopy, DFT, transport, magnetotransport, and magnetic measurements. Our results reveal a clear crossover from an electronically tuned regime at low Sb content to a disorder-dominated regime at higher Sb-substitution levels. For $x \leq 0.3$, superconductivity is only weakly suppressed from $T_c^{onset} \approx 48$ to $\approx 44$ K and is accompanied by a more homogeneous superconducting response, as evidenced by a narrower transition width and the disappearance of double-step features in magnetization curves. Enhanced vortex activation energies indicate an increased energy barrier for vortex motion and improved vortex pinning. Raman spectroscopy and DFT calculations confirm modifications of the Fe–pnictogen environment while maintaining robust superconductivity at low Sb concentrations. On the other hand, for $0.3 < x \leq 0.6$, superconductivity is rapidly suppressed, accompanied by a pronounced increase in residual resistivity ($\rho_0$), a dramatic reduction in the *RRR*, and broadening of the superconducting transition, reflecting the increasing importance of disorder-induced scattering and phase segregation. Quantitative phase analysis further reveals a progressive increase in Fe–Sb and Pr–Sb–O secondary phases, which contribute to the degradation of the superconducting properties at higher Sb concentrations. Although Sb substitution enhances the vortex activation energy and modifies the vortex dynamics, the global critical current density remains low, primarily due to degraded intergranular connectivity associated with secondary-phase formation. Overall, Sb substitution introduces non-magnetic disorder that initially preserves superconductivity while improving vortex pinning, but ultimately suppresses superconductivity at higher concentrations through a combination of electronic-structure modifications and disorder-induced scattering. The observed correlation between lattice expansion, phonon softening, residual scattering, vortex dynamics, and superconducting properties provides a unified framework for understanding the interplay between structural modification, disorder, and superconductivity in isovalently substituted 1111 iron pnictides.

## CRediT authorship contribution statement

**Priya Singh:** Writing – review & editing, Writing – original draft, Investigation, Formal analysis, Validation, Data curation. **Konrad Kwatek:** Formal analysis, Investigation, Resources, Data curation, Writing – review & editing. **Tatiana Zajarniuk:** Data curation, Investigation, Resources, Writing – review & editing. **Taras Palasyuk:** Writing – review & editing, Formal analysis, Investigation, Data curation. **Cezariusz Jastrzębski:** Writing – review & editing, Resources, Data curation. **Michał Wierzbicki:** Writing–review & editing, Investigation, Resources, Methodology. **Tomasz Cetner:** Data curation, Writing – review & editing. **Andrzej Szewczyk:** Writing–review & editing, Resources, Data curation. **Shiv J. Singh:** Writing–review & editing, Writing – original draft, Visualization, Validation, Supervision, Software, Resources, Methodology, Investigation, Funding acquisition, Formal analysis, Conceptualization.

## Declaration of competing interest

The authors declare that they have no known competing financial interests or personal relationships that could have appeared to influence the work reported in this paper.

## Data availability

The raw/processed data required to reproduce these findings cannot be shared at this time due to technical or time limitations. Data are available upon request to the corresponding author.

**Acknowledgments:**

This work was funded by SONATA-BIS 11 project (Registration number: 2021/42/E/ST5/00262) and the Weave-UNISONO project (2025/07/Y/ST5/00116) sponsored by National Science Centre (NCN), Poland. SJS acknowledges financial support from National Science Centre (NCN), Poland through research Project number: 2021/42/E/ST5/00262 and 2025/07/Y/ST5/00116.

**Figure 1:** **(a)** Powder X-ray diffraction (XRD) patterns of $PrFeAs_{1-x}Sb_xF_{0.7}O_{0.3}$ for $x$ = 0, 0.05, 0.15, 0.20, 0.30, 0.40, and 0.60. Data for higher Sb concentrations ($x$ = 0.8 and 1.0) are provided in Supplementary Figure S1. **(b)** Representative Rietveld refinement for $x$ = 0.15, where the experimental data are shown by open circles, the fitted profile by the red line, and the difference curve at the bottom; vertical ticks indicate Bragg reflection positions of the main PrFeAs(O,F) phase. **(c)** Evolution of the lattice parameter $c$ as a function of Sb content ($x$), compared with literature data from Azam et al. [12]. The inset shows the corresponding variation of the unit cell volume ($V$).

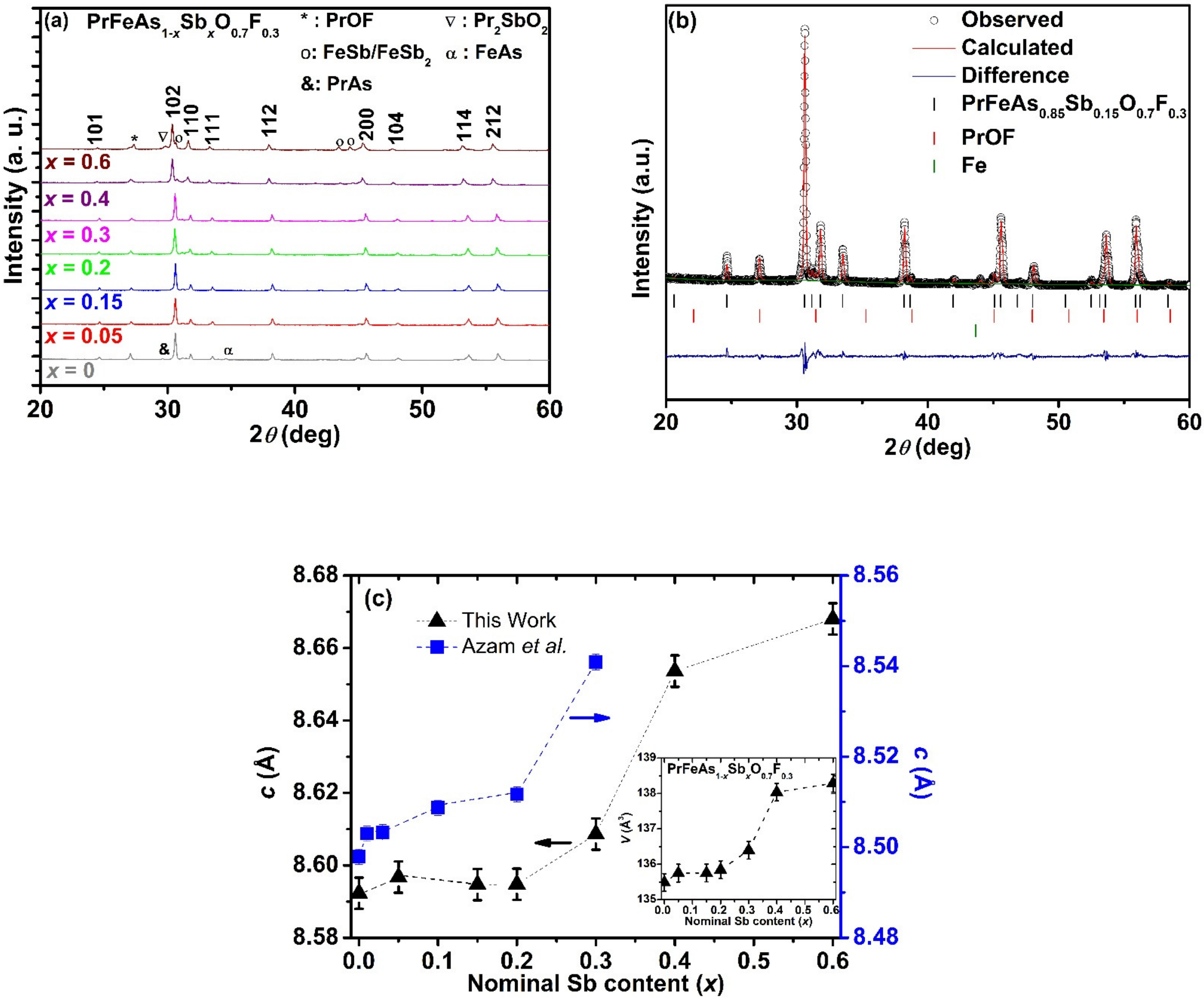

**Figure 2:** Raman spectroscopy study of Sb-doped $PrFe_{1-x}S_xAsO_{0.7}F_{0.3}$ series. (a) Representative Raman spectra of the pristine compound ($x$ = 0) and doped materials with antimony content 30 % ($x$ = 0.3) and 60% ($x$ = 0.6). Spectra are vertically offset for clarity. Assignment of detected signals related to lattice -vibrations is shown for the spectrum of the pristine compound. Deconvolution of experimental spectra with Lorentz function is shown as green lines. Total fits of Lorentz model to experimental data are shown by red lines. Vertical black lines are a guide to the eye. **(b)** Evolution of phonon peak positions as a function of Sb content ($X_{Sb}$). Symbols represent experimentally extracted peak positions for Pr, pnictogen (As/Sb), and Fe modes; error bars are smaller than the symbol size. Dashed lines are guides to the eye. The inset shows DFT-calculated phonon frequencies for the parent (PrFeAsO) and the hypothetical fully substituted (PrFeSbO) compounds, highlighting the effect of Sb substitution on lattice dynamics.

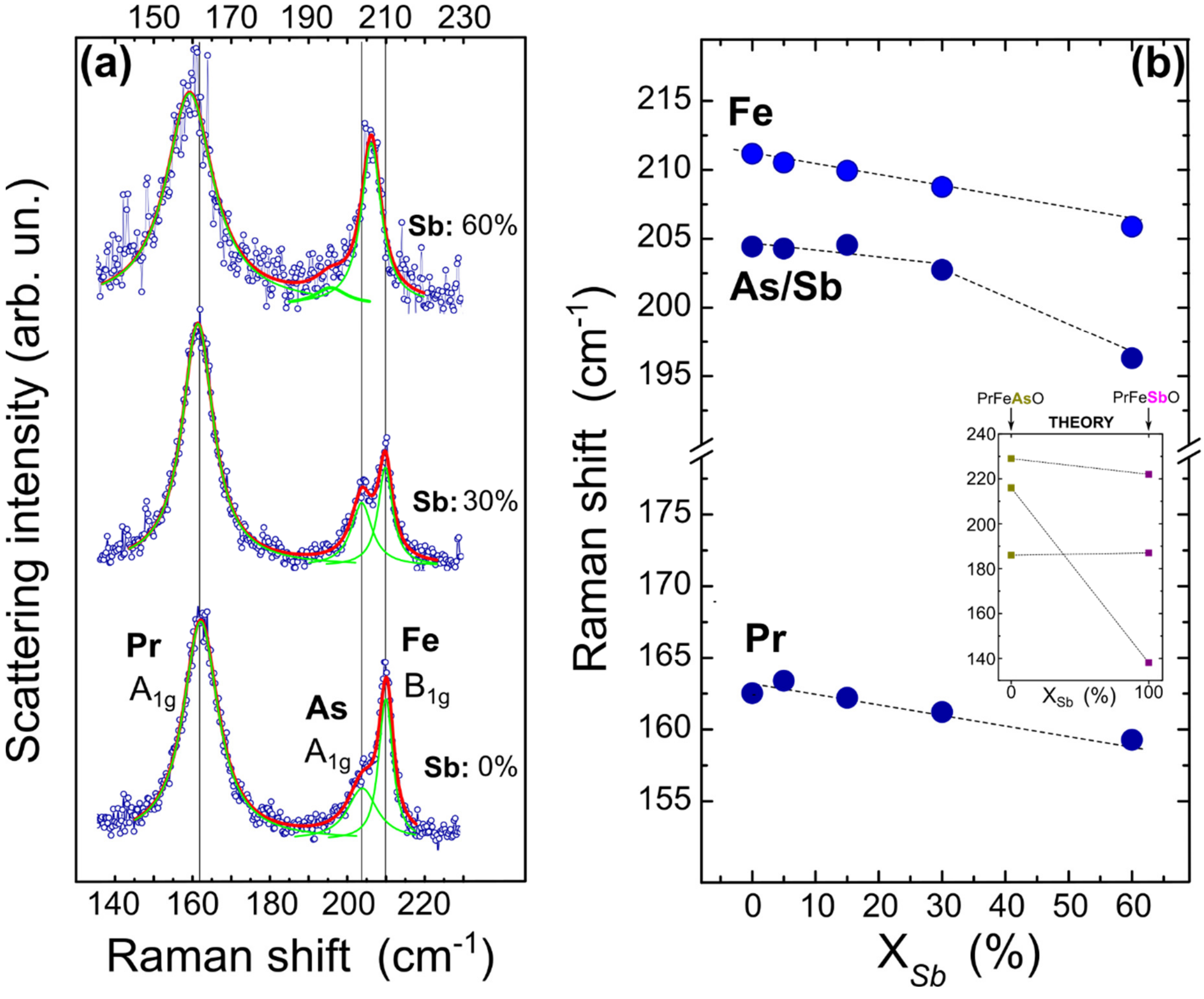

**Figure 3:** The temperature dependence of electrical resistivity for $PrFeAs_{1-x}Sb_xF_{0.7}O_{0.3}$ at zero magnetic field. **(a)** Resistivity as a function of temperature in the range 7–300 K for $x$ = 0, 0.05, 0.15, 0.20, 0.30, 0.40, and 0.60. **(b)** Expanded view of the low-temperature region (7–60 K), highlighting the superconducting transitions.

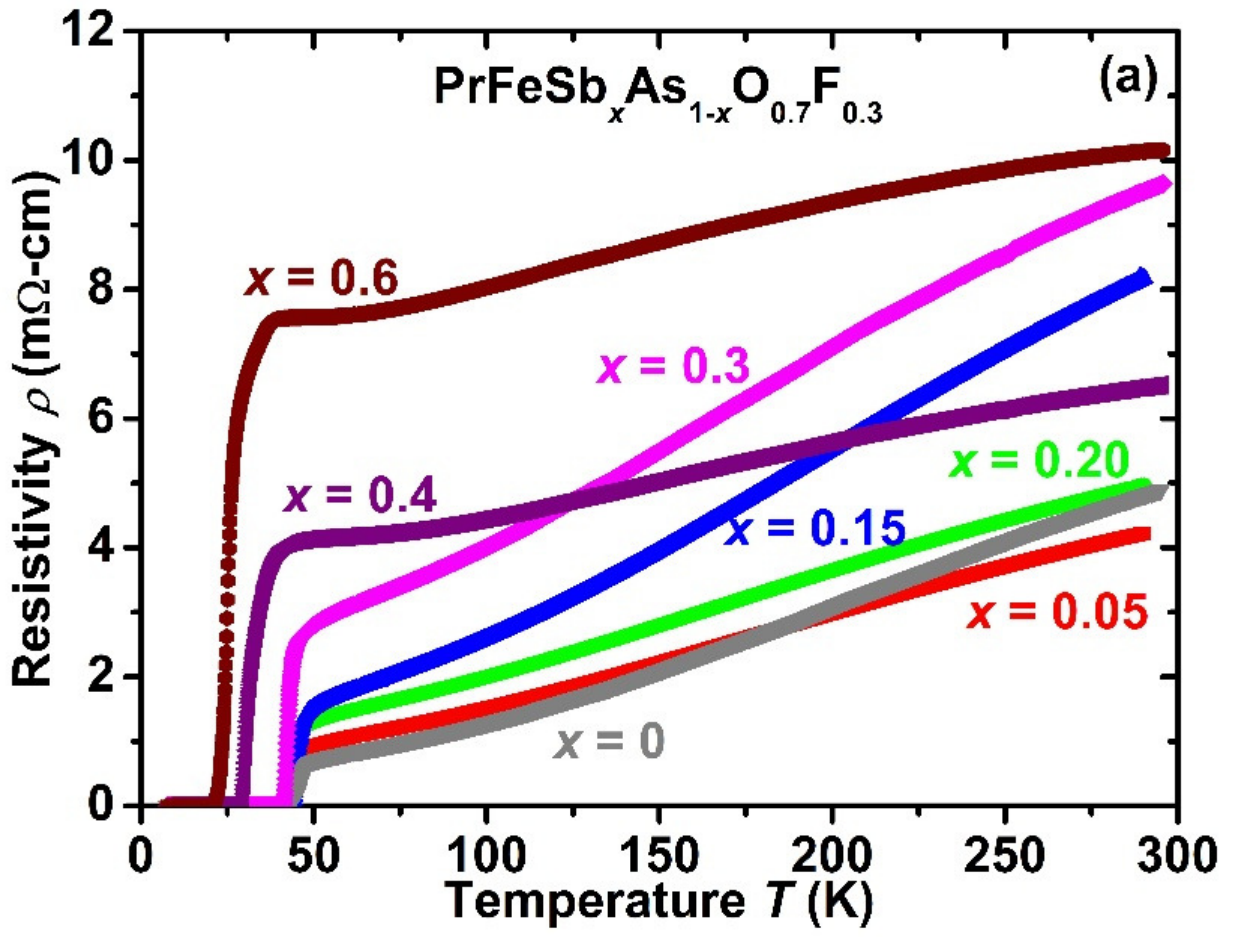

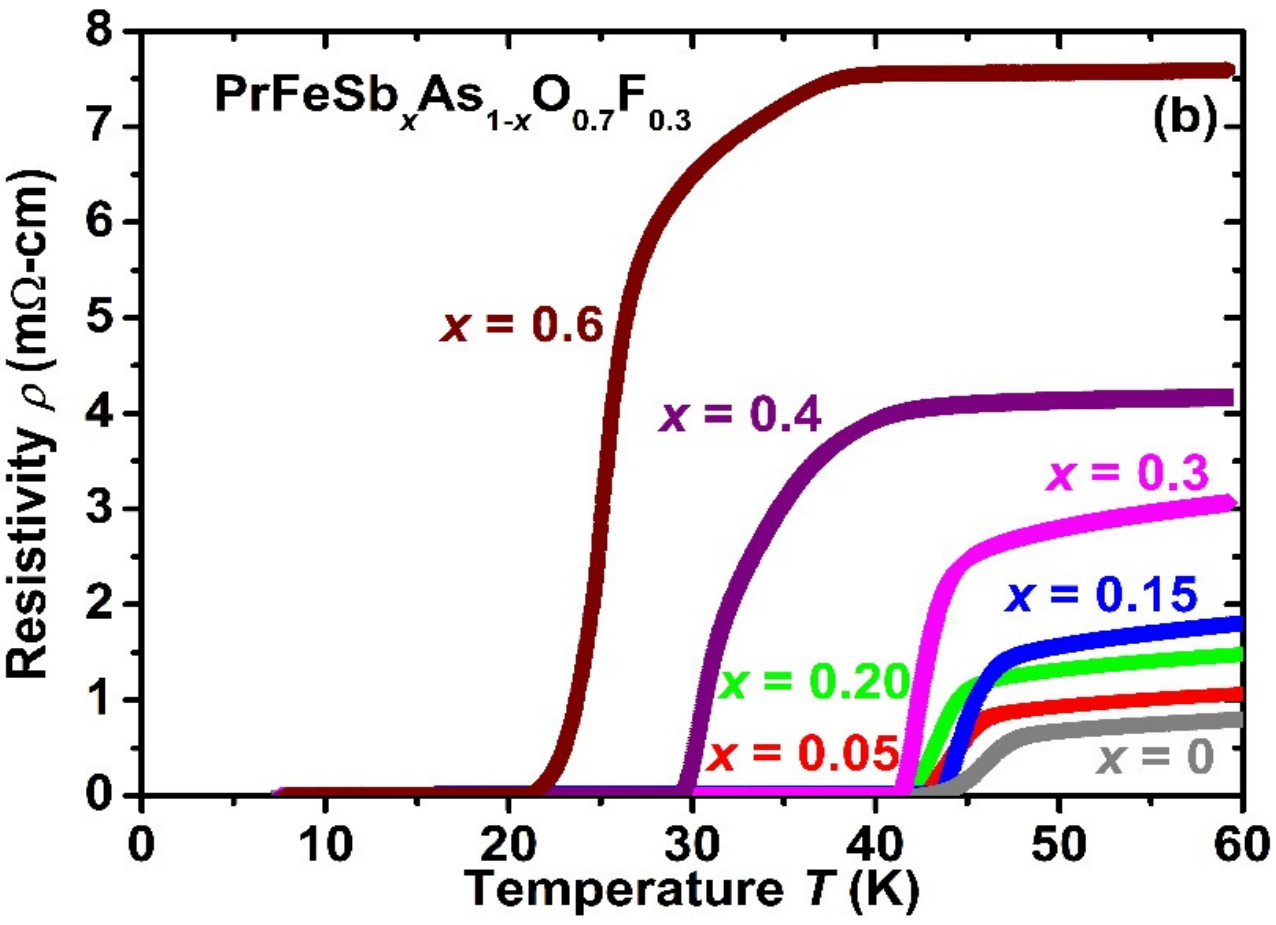

**Figure 4:** Magneto-transport properties of Sb-substituted $PrFeAs_{1-x}Sb_xF_{0.7}O_{0.3}$. **(a)** The temperature dependence of normalized resistivity for $x$ = 0.15 measured under applied magnetic fields from 0 to 9 T (normalized at 50 K) in the range 5–60 K, showing field-induced broadening of the superconducting transition. **(b)** Corresponding data for $x$ = 0.30. **(c)** Temperature dependence of the upper critical field $H_{c2}$ and irreversibility field ($H_{irr}$) for $x$ = 0, 0.15, and 0.30, determined using 90% and 10% resistivity criteria, respectively.

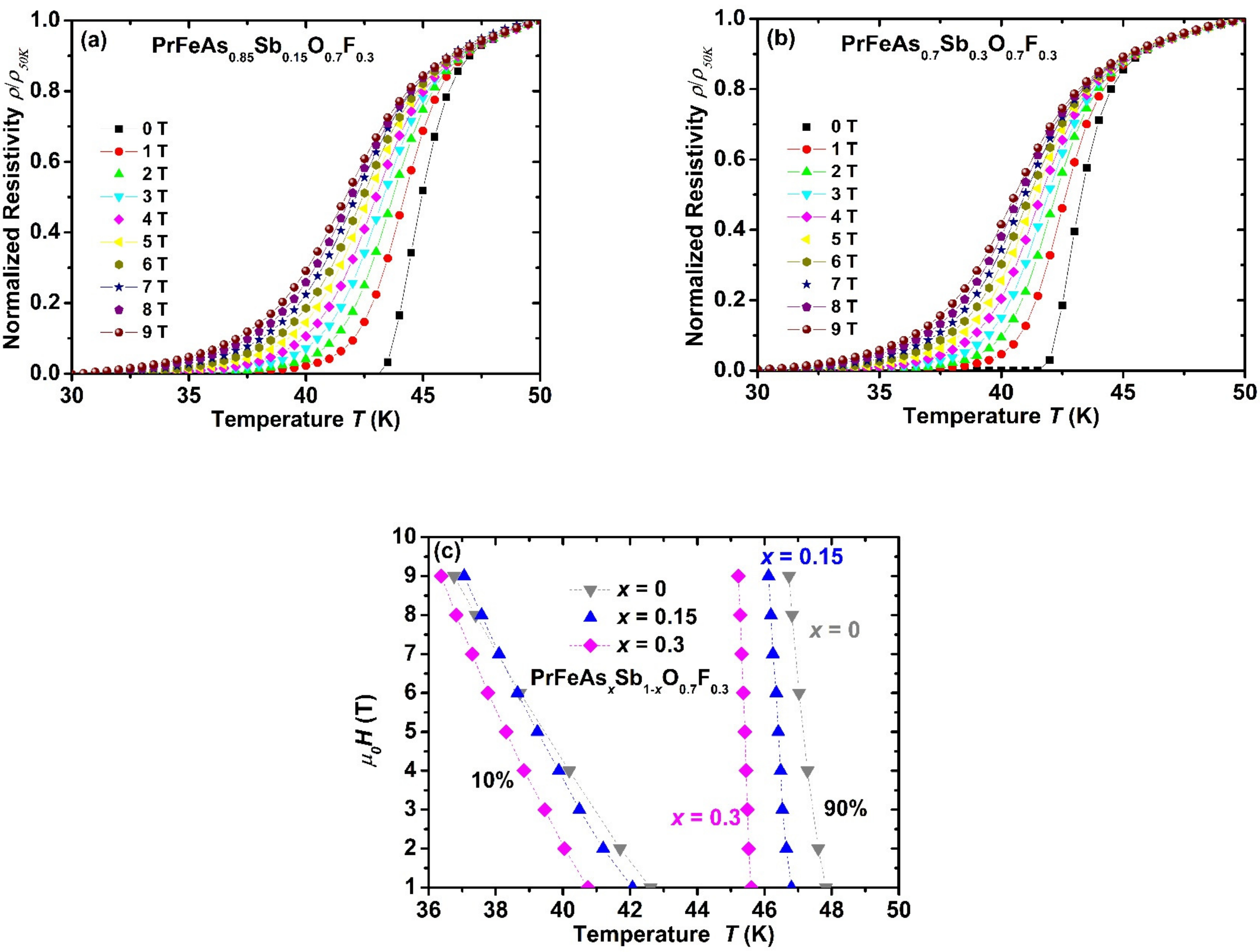

**Figure 5:** Thermally activated flux flow (TAFF) analysis of Sb-substituted $PrFeAs_{1-x}Sb_xF_{0.7}O_{0.3}$. Arrhenius plots of $\log(\rho/\rho_{50K})$ versus inverse temperature ($1/T$) for $x = 0.15$ **(a)** and $x = 0.30$ **(b)**, measured under applied magnetic fields from 0 to 9 T and normalized at 50 K. The linear regions correspond to TAFF behavior, from which the activation energy ($U_0$) is extracted. **(c)** Magnetic-field dependence of the activation energy $U_0$ for $x$ = 0, 0.15, and 0.30, following a power-law relation $U_0 \sim H^{-\eta}$.

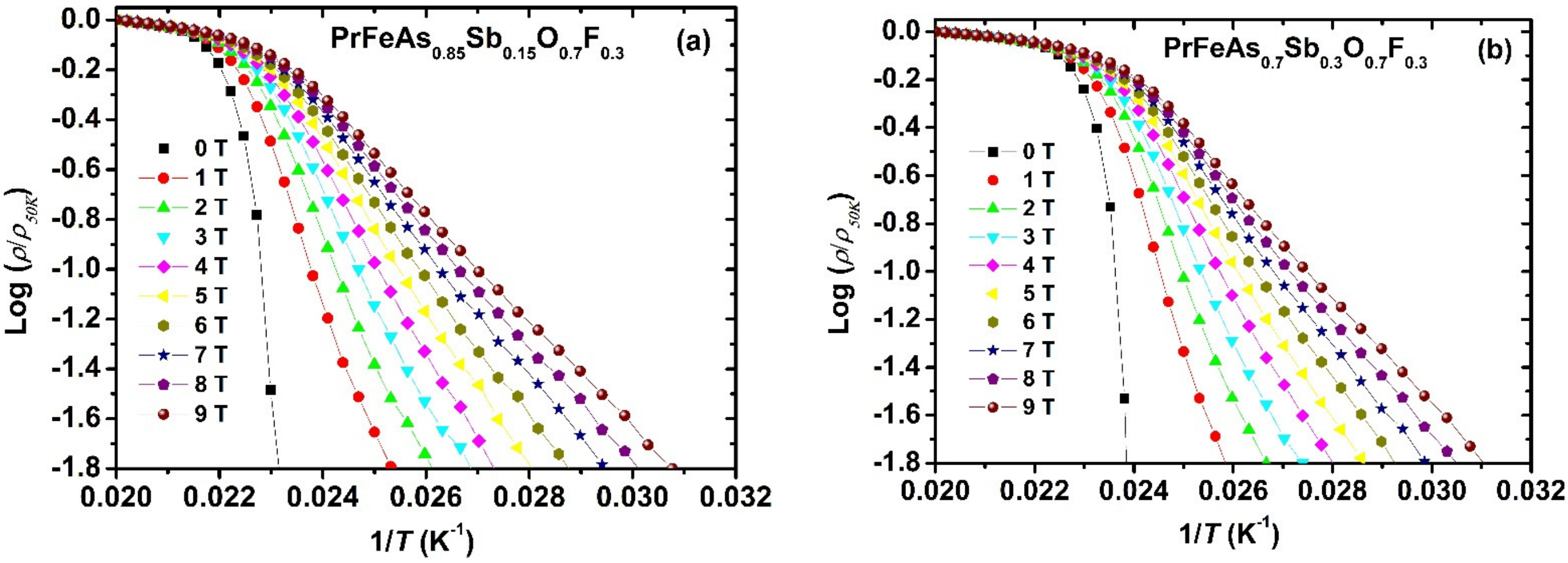

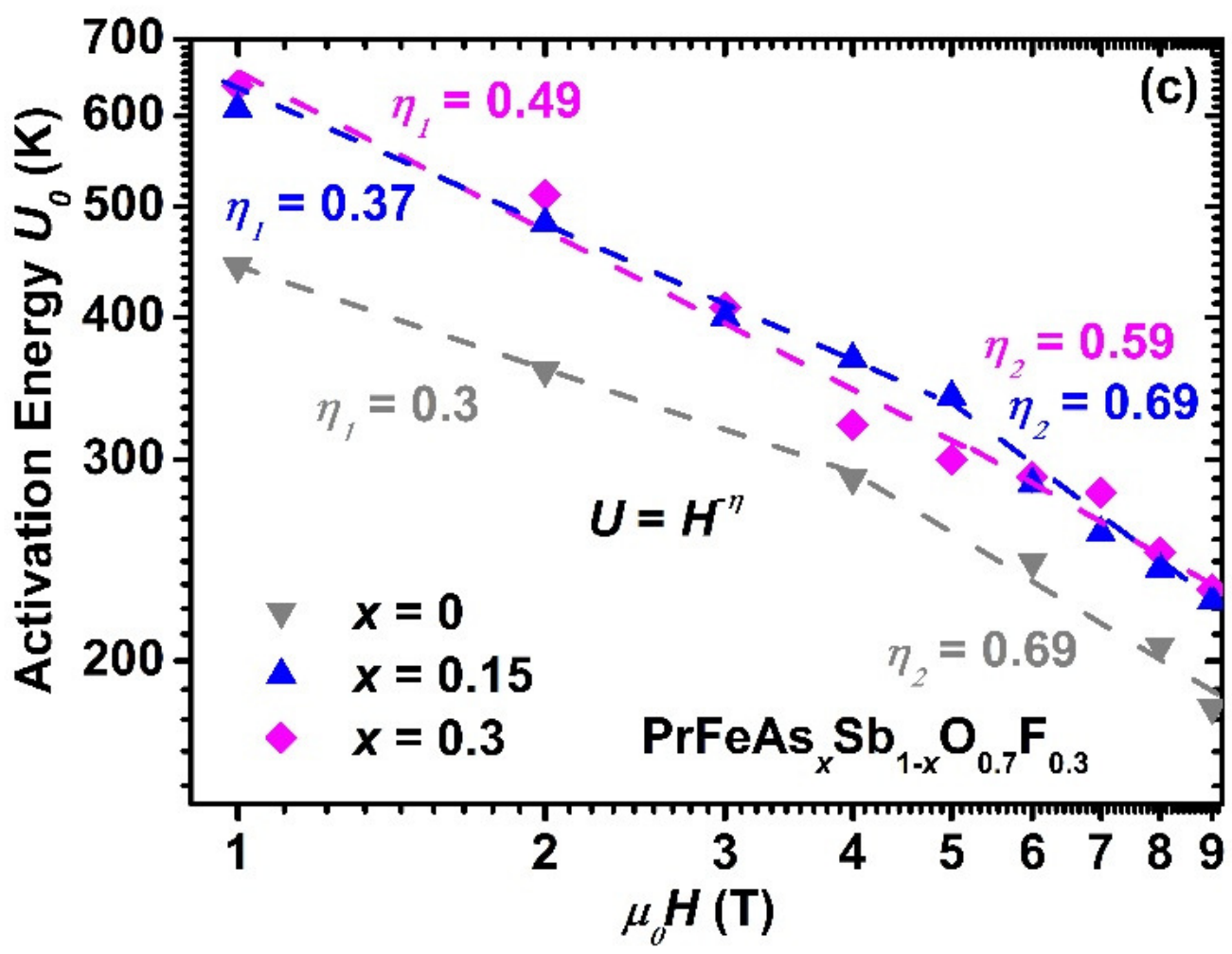

**Figure 6:** Magnetic properties of Sb-substituted $PrFeAs_{1-x}Sb_xF_{0.7}O_{0.3}$. **(a)** Temperature dependence of normalized magnetization ($M/M_{5K}$) for $x$ = 0, 0.15, 0.30, 0.40, and 0.60, measured under an applied magnetic field of 20 Oe in zero-field-cooled (ZFC) and field-cooled (FC) modes. **(b)** Magnetic-field dependence of the critical current density ($J_c$) at $T$ = 5 K for $x$ = 0, 0.15, and 0.30, estimated from magnetization loops using the Bean model.

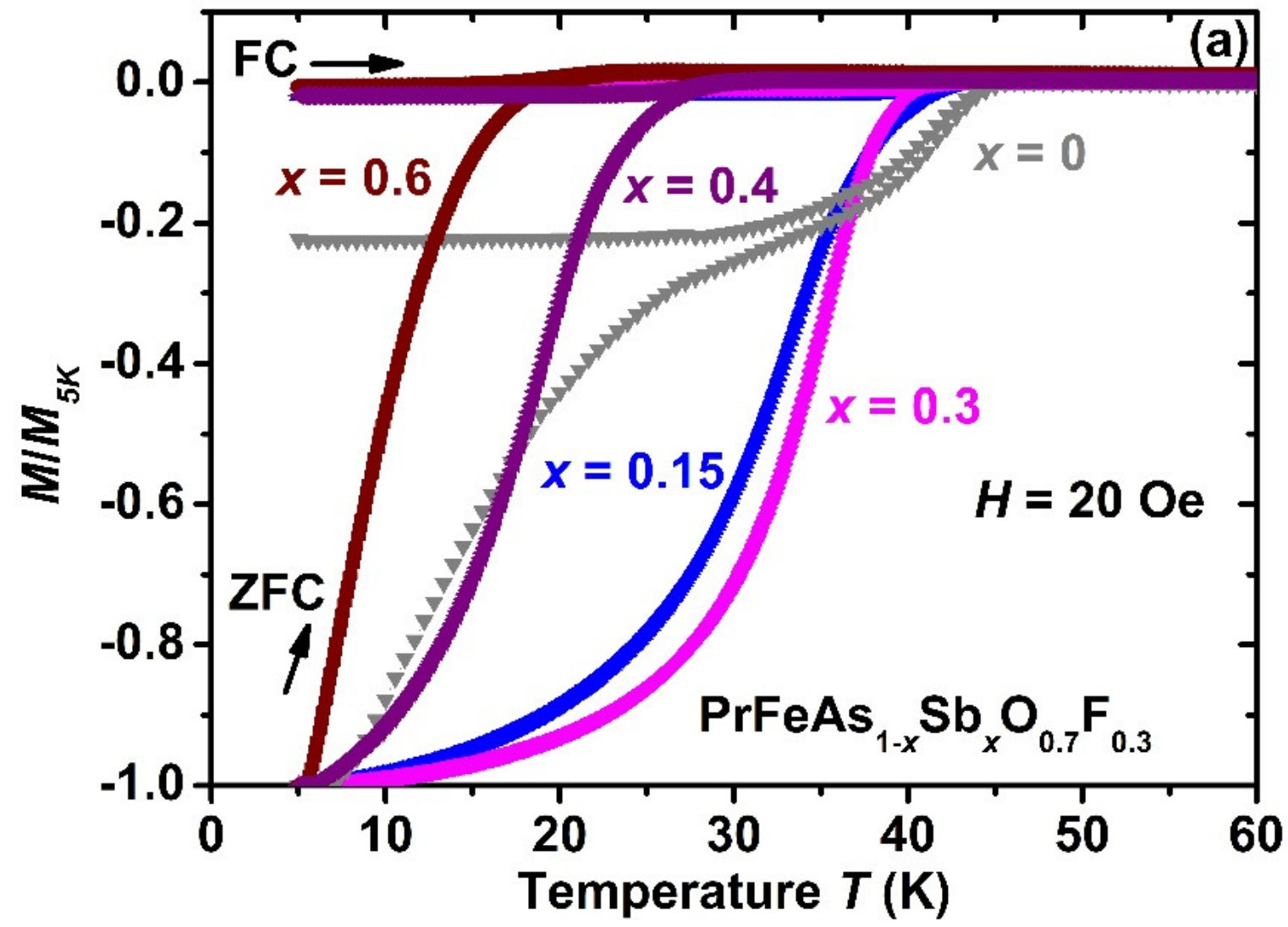

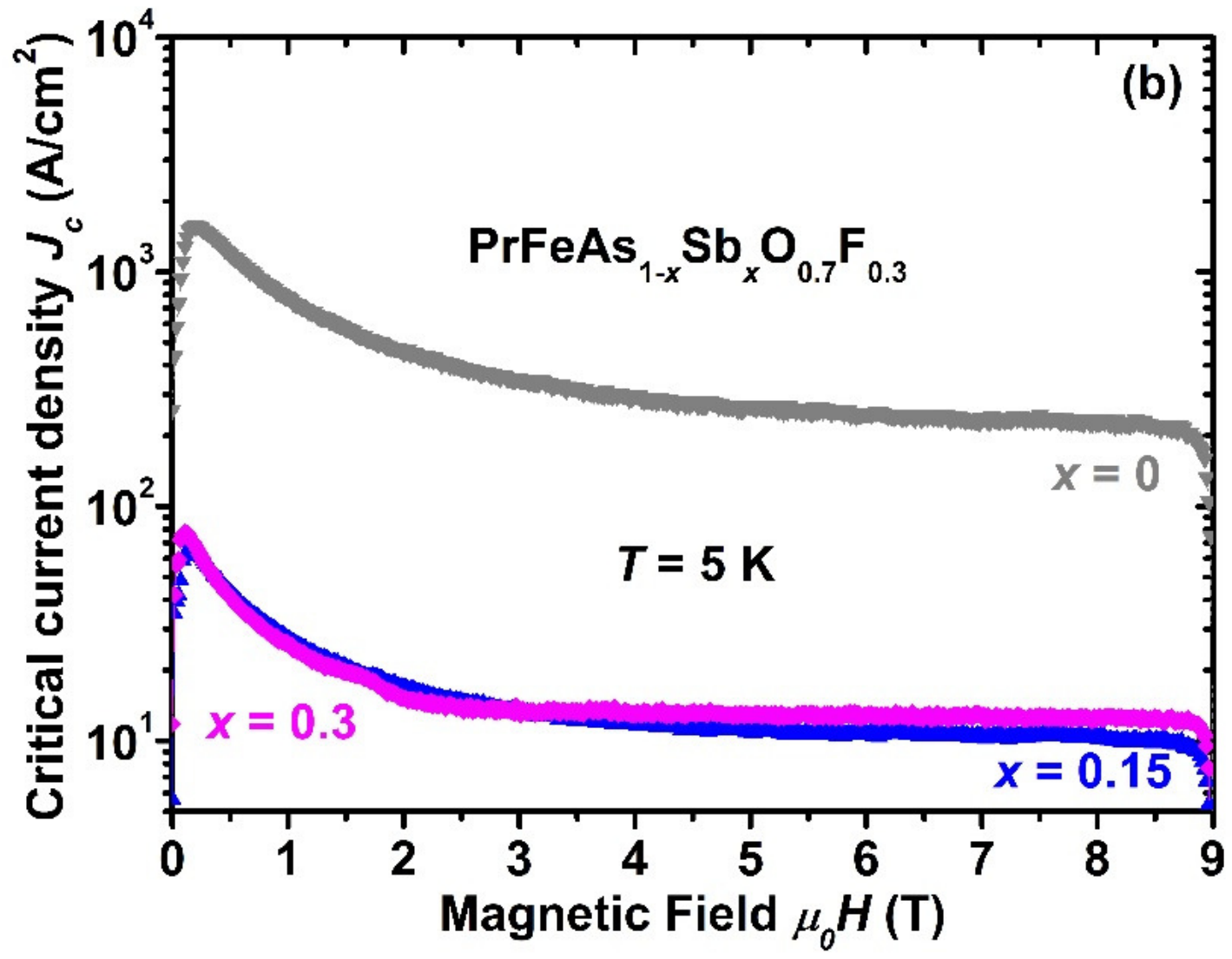

**Figure 7:** Evolution of superconducting and normal state parameters for $PrFeAs_{1-x}Sb_xF_{0.7}O_{0.3}$ ($0 \leq x \leq 0.6$). The variation of (**a**) superconducting transition temperature $T_c^{onset}$ obtained from the transport measurements; (**b**) the superconducting transition width $\Delta T$; (**c**) residual resistivity $\rho_0$ obtained from extrapolation of the normal-state resistivity to $T = 0$ K and **(d)** residual resistivity ratio $RRR = \rho_{300\,K}/\rho_{50\,K}$ and $RRR = \rho_{300K}/\rho_0$ as a function of Sb concentration ($x$). The observed trends are compared with previously reported data for Sb-doped $SmFeAs_{1-x}Sb_xO_{0.8}F_{0.2}$ (Azam *et al.*) [12]. In panel (d), the arrows indicate the corresponding $y$-axis for the conventional $RRR$ (= $\rho_{300\,K}/\rho_{50\,K}$), the modified $RRR$ (= $\rho_{300K}/\rho_0$), and the literature data from Azam et al. [12].

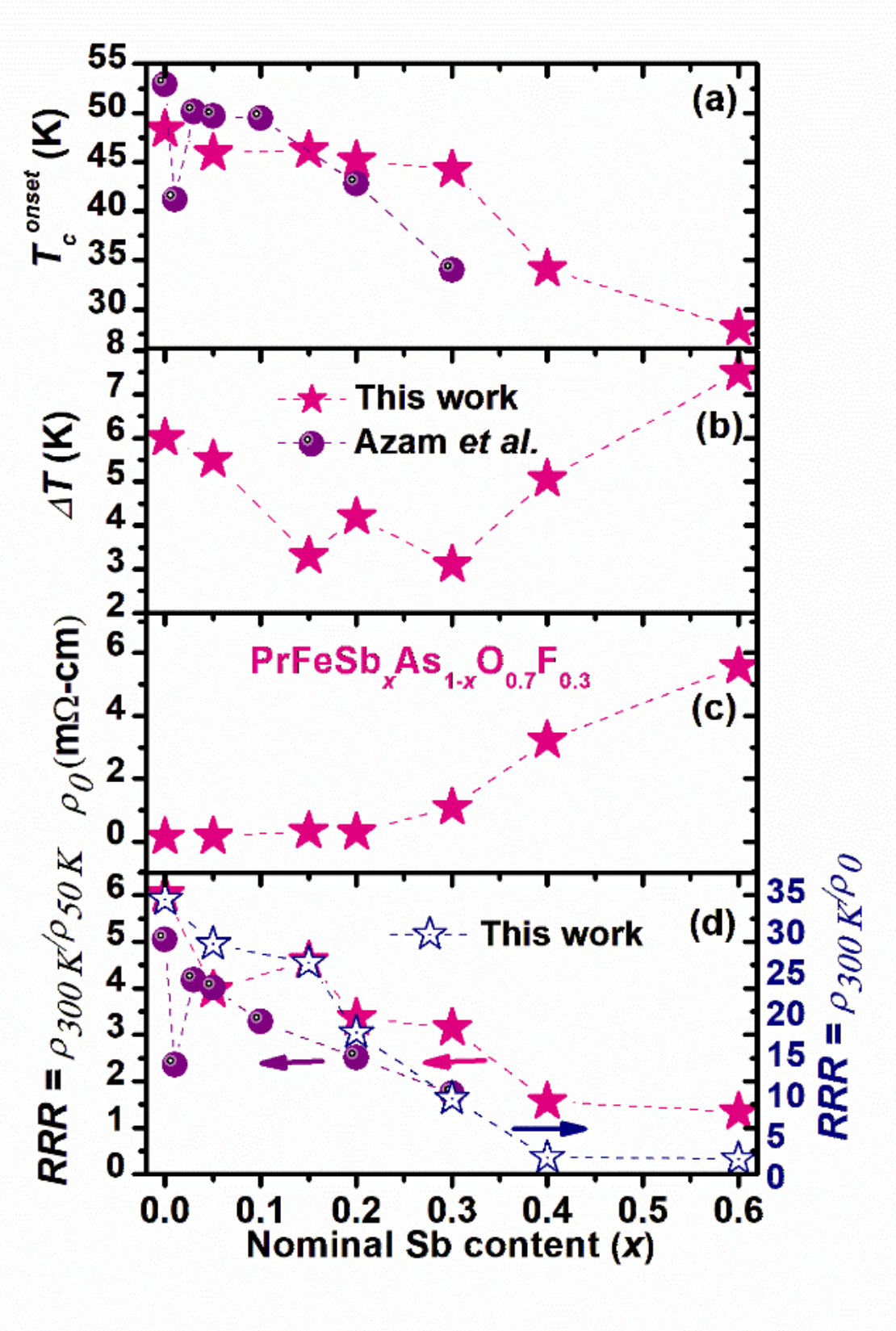